\documentclass[acmlarge]{acmart}

\usepackage{ulem}
\usepackage{multirow}
\usepackage{listings}
\usepackage{hyperref}

\newcommand{\anonymous}[2]{#1}  

\AtBeginDocument{%
  \providecommand\BibTeX{{%
    \normalfont B\kern-0.5em{\scshape i\kern-0.25em b}\kern-0.8em\TeX}}}

\setcopyright{acmcopyright}
\copyrightyear{2022}
\acmYear{2022}
\acmDOI{XXXXXXX.XXXXXXX}

\acmJournal{DTRAP}

\definecolor{codegreen}{rgb}{0,0.6,0}
\definecolor{blue}{rgb}{0.0, 0.0, 1.0}
\definecolor{bondiblue}{rgb}{0.0, 0.58, 0.71}
\definecolor{brinkpink}{rgb}{0.98, 0.38, 0.5}
\definecolor{brandeisblue}{rgb}{0.0, 0.44, 1.0}
\definecolor{codegray}{rgb}{0.5,0.5,0.5}
\definecolor{codepurple}{rgb}{0.58,0,0.82}
\definecolor{backcolour}{rgb}{0.95,0.95,0.92}
    
\lstdefinestyle{mystyle}{
    commentstyle=\color{blue},
    keywordstyle=\color{magenta},
    numberstyle=\tiny\color{codegray},
    stringstyle=\color{codepurple},
    basicstyle=\ttfamily\footnotesize,
    breakatwhitespace=false,         
    breaklines=true,                 
    captionpos=b,                    
    keepspaces=true,                 
    numbers=left,                    
    numbersep=5pt,                  
    showspaces=false,                
    showstringspaces=false,
    showtabs=false,                  
    tabsize=2,
    moredelim=[is][\underbar]{_}{_},
    keepspaces=true
}
\begin{document}

\title{Industrial Experience of Finding Cryptographic Vulnerabilities in Large-scale Codebases}

\author{Ya Xiao}
\authornote{The work was performed while the first author was at Oracle Labs as an intern.}
\affiliation{%
  \institution{Virginia Tech}
  \city{Blacksburg}
  \country{USA}}
\email{yax99@vt.edu}

\author{Yang Zhao}
\affiliation{%
  \institution{Oracle Labs}
  \city{Brisbane}
  \country{Australia}
}
\email{yang.yz.zhao@oracle.com}

\author{Nicholas Allen}
\affiliation{%
 \institution{Oracle Labs}
 \city{Brisbane}
 \country{Australia}
}
\email{nicholas.allen@oracle.com}

\author{Nathan Keynes}
\affiliation{%
 \institution{Oracle Labs}
 \city{Brisbane}
 \country{Australia}
}

\author{Danfeng (Daphne) Yao}
\affiliation{%
  \institution{Virginia Tech}
  \city{Blacksburg}
  \country{USA}}
\email{danfeng@vt.edu}

\author{Cristina Cifuentes}
\affiliation{%
 \institution{Oracle Labs}
 \city{Brisbane}
 \country{Australia}
}
\email{cristina.cifuentes@oracle.com}


\begin{abstract}
Enterprise environment often screens large-scale (millions of lines of code) codebases with static analysis tools to find bugs and vulnerabilities. Parfait is a static code analysis tool used in \anonymous{Oracle}{a large software company} to find security vulnerabilities in industrial codebases.  Recently, many studies show that there are complicated cryptographic vulnerabilities caused by misusing cryptographic APIs in Java\anonymous{\textsuperscript{TM}\footnote{Java is a registered trademark of Oracle and/or its affiliates.}}{}. In this paper, we describe how we realize a precise and scalable detection of these complicated
cryptographic vulnerabilities based on Parfait framework.  
The key challenge in the detection of cryptographic vulnerabilities is the high false alarm rate caused by  pseudo-influences. Pseudo-influences happen if security-irrelevant constants are used in constructing security-critical values. Static analysis is usually unable to distinguish them
from hard-coded constants that expose sensitive information.  We tackle this problem by specializing the backward dataflow analysis used in Parfait with refinement insights, an idea from the tool CryptoGuard~\cite{rahaman2019cryptoguard}.  
We evaluate our analyzer on a comprehensive Java cryptographic vulnerability benchmark and eleven large real-world applications. The results show that the Parfait-based cryptographic vulnerability detector can find real-world cryptographic vulnerabilities in large-scale
codebases with high true-positive rates and low runtime cost.
\end{abstract}

\ccsdesc[500]{Security and privacy~Software and application security}

\keywords{Cryptography, misuse, API, static analysis}

\maketitle

\section{Introduction}
To guarantee the security of large projects, companies usually deploy various bug checking tools in the  development process.  Parfait~\cite{cifuentes2008parfait} is such a static code analysis tool designed for large-scale codebases to find security and quality defects in C/C++, Java, Python, and PL/SQL languages. In particular, Parfait focuses on defects from the lists of CWE Top 25~\cite{cwetop25} and OWASP Top 10~\cite{owasptop10}.
Cryptographic vulnerabilities caused by misusing Java Cryptographic APIs are getting more and more attention~\cite{acar2016you,meng2018secure,georgiev2012most,egele2013empirical,zuo2019does}. A survey shows that cryptographic API misuses dominate the cryptographic vulnerabilities, accounting for 83\% in ``cryptography issues'' category of the Common Vulnerabilities and Exposures (CVE) database~\cite{lazar2014does}. Cryptographic failure has been recognized as the second risk in OWASP Top 10 for 2021~\cite{owasptop10}. Java provides basic cryptographic objects (e.g., \lstinline|Cipher|, \lstinline|MessageDigest|) in Java Cryptography Architecture (JCA) and Java Cryptography Extension (JCE) libraries. Due to complex documentation and the lack of security expertise, developers may not know how to use these APIs correctly~\cite{nadi2016jumping,acar2017comparing}.  Parfait supports the detection of simple cryptographic vulnerabilities, such as using broken Cipher or Hash algorithms. 
However, many studies show that cryptographic API misuses are more complicated and involve more security rules~\cite{fahl2012eve,egele2013empirical,nguyen2017stitch,meng2018secure,DBLP:conf/ccs/BosuLYW17,DBLP:journals/tdsc/TianYRTP20,patnaik2019usability}. 

Software developers struggle to understand and comply with the implicit and explicit requirements of using cryptographic APIs securely. Violating these requirements may cause various vulnerabilities including exposing sensitive information, bypassing necessary authentication, etc.  Egele et al.~\cite{egele2013empirical} identified  six types of cryptographic API misuses that violate different security rules. Nguyen et al.~\cite{nguyen2017stitch} showed
   thirteen security pitfalls common in Android development and nine of them are Java cryptographic API misuses. Recently, Rahaman et al.~\cite{rahaman2019cryptoguard} summarized sixteen common types of cryptographic API misuses in Java and developed the CryptoGuard tool to detect them. It relies on backward and forward program slicing and introduces several refinement insights to achieve high precision and scalability in large projects. 
  
  We extended Parfait with a precise and scalable dataflow analysis to detect Java cryptographic API misuse vulnerabilities. Parfait offers a proprietary compilation process to transform Java source code into the low level virtual machine
 (LLVM) intermediate representation (IR). 
   In particular, we need to develop a precise and scalable cryptographic API misuse detection on top of LLVM IR with Parfait's supports. In this work, we identify eleven cryptographic vulnerability types (see Table~\ref{tab:crypto_api_usage}) that can be mapped to backward dataflow analysis problems. By monitoring their different vulnerable usages, we designed corresponding alarm criteria. For example, the alarm criterion for the vulnerability ``Use of a Broken or Risky Cryptographic Algorithm'' is a constant matching given weak algorithm names (e.g., "DES") and the alarm criterion for the vulnerability ``Use of Password Hash With Insufficient Computational Effort'' is an iteration count number less than 1000. 
 
Cryptographic vulnerabilities are difficult to identify precisely.  
 Most of these vulnerabilities are caused by assigning inappropriate values (e.g., hard-coded values) to sensitive information (e.g., keys, passwords) that are required to be secret or unpredictable. To detect them, the backward dataflow analysis is used to trace all the sources influencing these security-critical variables in a program. Sources that are constants are treated as hard-coded values and they may be reported as vulnerabilities. However, this technique can cause many false alarms. There are many cases that involve constants in constructing a non-constant value~\cite{afrose2019cryptoapi}. For example, a constant string can represent a file location where the secret key is loaded. Those constants that do not impact security are called \textit{pseudo-influences} in the work of CryptoGuard~\cite{rahaman2019cryptoguard}, which has identified five types of pseudo-influences (e.g., state indicator) and refinement insights to reduce them. In our work, these refinement insights are further adjusted to improve detection precision.


 We built our cryptographic vulnerability detection using Parfait framework with its many built-in program analysis techniques.  In particular, we specialize the IFDS analysis, which is a dataflow analysis framework for interprocedural, finite, distributive subset (IFDS) problems~\cite{reps1995precise}, for cryptographic vulnerability detection. It allows program analysis designers to configure API methods as taint sources or sinks, and then check whether there is a dataflow from a source to a sink. In this work, we first identify the sensitive variables by setting eighteen error-prone Java cryptographic API methods (see Table~\ref{tab:crypto_api_usage}) as sinks. Because ordinary taint analysis does not track constants, we further modify the taint analysis to be capable of tracking all constant sources.  Moreover, we refine the taint analysis by eliminating tracing the pseudo-influences identified by the refinement rules of CryptoGuard. This refinement significantly reduces the false alarms and improves efficiency by eliminating unnecessary dataflows. Finally, we improve the scalability by leveraging Parfait's layered framework to break down the interprocedural analysis into method-level pieces and schedule them adaptively.

Our contributions are summarized as follows:
\begin{itemize}
    \item We realized the detection for complex Java cryptographic vulnerabilities in \anonymous{Oracle's}{} Parfait static analysis platform. Specifically, we implemented analyses for eleven CWE types caused by misusing eighteen associated Java cryptographic API methods. The detection relies on a backward inter-procedural, flow-, context-, field-sensitivity dataflow analysis with Parfait and LLVM supports. We designed different alarm criteria for identifying these cryptographic vulnerabilities.

    \item We specialized the backward IFDS taint analysis provided by Parfait to overcome the precision challenge caused by pseudo-influences, security-irrelevant constants used in constructing security-critical values. Inspired by the refinement insights in CryptoGuard~\cite{rahaman2019cryptoguard},  we defined the refinement rules in the form of IFDS dataflow analysis. Significantly, the refined analysis not only reduces false alarms but also improves scalability. 

    \item  We evaluated the precision and scalability of Parfait cryptographic vulnerability detection on a comprehensive cryptographic vulnerability benchmark CryptoAPI-Bench~\cite{afrose2019cryptoapi} and several large-scale industrial applications. The results demonstrate that our detection achieves a high precision (86.62\%) and recall (98.40\%) overall. The precision excluding the path-sensitivity test cases reaches 100\%. Parfait-based cryptographic vulnerability detection achieves 100\% precision on the eleven large-scale applications. The runtime for analyzing the codebases with sizes from 2K to 1321K lines of code ranges from 2 seconds to 36 minutes, with the majority of the codebases analyzed within ten minutes.  We further show some noteworthy examples to help readers better understand the practices.
    
\end{itemize}

 In summary, we have  developed a precise and scalable analysis to detect cryptographic vulnerabilities. Our work incorporates the false positive reduction refinements of CryptoGuard, the scalable framework of Parfait, and the IFDS analysis on top of LLVM IR. The evaluation results show that our tool works well in an industrial setting.  
\section{Background}

This section describes the Java cryptographic API misuses that are the targets of our detection and provides background of CryptoGuard and the \anonymous{Oracle}{industrial} Parfait static analysis framework.

\subsection{Java Cryptographic API misuses}

Table~\ref{tab:crypto_api_usage} lists the targeted Java cryptographic API misuses from the developer's perspective, with the involved API classes, methods, and the vulnerable usages of them. We summarize these Java Cryptographic API misuses that can be detected by backward dataflow analysis from the existing studies~\cite{rahaman2019cryptoguard,egele2013empirical,nguyen2017stitch}. Compared with CryptoGuard, it does not cover a few vulenrability types that require combining forward analysis with backward analysis to detect. 

\begin{table*}[!htb]
\centering
\caption{Error-prone Java Cryptographic APIs covered by Parfait's cryptographic API misuses detection and the eleven involved vulnerability types in CWE. The severity information is from CryptoGuard~\cite{rahaman2019cryptoguard}.}
\begin{scriptsize}
\begin{tabular}{clllcl}
\hline
  & \textbf{Class}                   & \textbf{Method Names}                 & \multicolumn{1}{c}{\textbf{Vulnerable Usage}}    &\textbf{Severity} &  \multicolumn{1}{c}{\textbf{CWE}}                     \\ \hline
1 & Random                           & constructor                            & used in cryptography operations          & M          & 338: Use of Cryptographically Weak PRNG                         \\ \hline
2 &\multirow{2}{*}{SecureRandom}     & constructor                      & \multirow{2}{*}{pass static or predictable seed}   &\multirow{2}{*}{M}   & \multirow{2}{*}{337: Predictable Seed in PRNG}                           \\
3 &                                  & setSeed                               &                                                  &                                                                            \\ \hline
4 & MessageDigest                    & getInstance                           & pass weak algorithm                     & H          & 328: Reversible One-Way Hash                                         \\ \hline
5 & \multirow{2}{*}{Cipher}          & \multirow{2}{*}{getInstance}          & pass weak algorithm                    &\multirow{2}{*}{L}           & \multirow{2}{*}{327: Use of a Broken or Risky Cryptographic Algorithm}    \\
6 &                                  &                                       & pass ECB mode for block ciphers                   &                                                                            \\ \hline
7 &\multirow{4}{*}{KeyStore}         & load                                  & \multirow{4}{*}{pass hard-coded password}      &  \multirow{4}{*}{H}     & \multirow{4}{*}{259: Use of Hard-coded Password}                      \\
8 &                                  & store                                 &                                                  &                                                                            \\
9 &                                  & setKeyEntry                           &                                                  &                                                                            \\
10&                                  & getKey                                &                                                  &                                                                            \\ \hline
11&SecretKeySpec                     & constructor                     & pass hard-coded key materials            & H          & 321: Use of Hard-coded Cryptographic Key                                 \\ \hline

12&\multirow{3}{*}{PBEKeySpec}       & \multirow{3}{*}{constructor}       & pass hard-coded password                    & H       & 259: Use of Hard-coded Password                                         \\
13&                                  &                                       & pass static or predictable salt                &M    & 760: Use of a One-Way Hash with a Predictable Salt                      \\
14&                                  &                                       & pass iteration \textless{}1000                 &L      & 916: Use of Password Hash With Insufficient Computational Effort       \\ \hline
15&\multirow{2}{*}{PBEParameterSpec} & \multirow{2}{*}{constructor} & pass static or predictable salt         &M         & 760: Use of a One-Way Hash with a Predictable Salt                         \\
16&                                  &                                       & pass iteration \textless{}1000       &M            & 916: Use of Password Hash With Insufficient Computational Effort           \\ \hline
17&IvParameterSpec                   & constructor                   & pass static or predictable IV              & M         & 329: Not Using a Random IV with CBC Mode                             \\ \hline
18&\multirow{3}{*}{TrustManager}     & checkClientTrusted                    & override to skip validation         &\multirow{3}{*}{H}              & \multirow{3}{*}{303: Incorrect Implementation of Authentication Algorithm}\\
19&                                  & checkServerTrusted                    & override to skip validation                      &                                                                            \\
20&                                  & getAcceptedIssuers                    & override to return null                                             &                                                                \\ \hline
21&HostnameVerifier                  & verify                                & override to always return True              &H     & 303: Incorrect Implementation of Authentication Algorithm                  \\ \hline
22&SSLSocketFactory                  & createSocket                         & miss hostname verification                  &H       & 304: Missing Critical Step in Authentication                            \\ \hline
\end{tabular}
\end{scriptsize}
\label{tab:crypto_api_usage}
\end{table*}

\noindent
The involved error-prone Java classes include:

\smallskip
\noindent
\verb|SecureRandom| \textit{Class}. Any nonce used in cryptography operations should be generated with \verb|SecureRandom| instead of \verb|Random|.  Furthermore, setting a static or predictable seed via the constructors or \verb|setSeed| methods\footnote{This API has two different method signatures (setSeed(long seed) and setSeed(byte[] seed)), we skip them for simplicity.} is also considered vulnerable. 

\smallskip
\noindent
\verb|MessageDigest| \textit{Class}. Passing a broken hash algorithm (e.g., MD5) to \verb|getInstance| method of \verb|MessageDigest| class is vulnerable.

\smallskip
\noindent
\verb|Cipher| \textit{Class}. The method \verb|getInstance| of \verb|Cipher| class is error-prone of using broken ciphers or insecure mode. The specific vulnerable usages include 1) passing a weak cipher algorithm (e.g., \verb|"DES"|); 2) specifying \verb|"ECB"| mode for a block cipher (e.g., \verb|"AES/ECB/NoPadding"|); 3) a block cipher without explicitly specifying a mode (e.g., \verb|"AES"|) because the vulnerable mode \verb|ECB| is used by default.


\smallskip
\noindent 
\verb|KeyStore| and \textit{Key Specification Classes}. Many API methods of \verb|KeyStore| and various key specification classes (e.g., \verb|SecretKeySpec|, \verb|PBEKeySpec|) accept secrets (e.g., passwords, key materials) by passing them through the method arguments. Any method call accepting a hard-coded or predictable secret is vulnerable.

\smallskip
\noindent
\textit{Algorithm Parameter Classes}. Algorithm parameter classes, such as \verb|IvParameterSpec| and \verb|PBEParameterSpec|, work with the initial vector (IV), salt, and PBE iteration count. IVs and salts that are static or predictable can cause vulnerabilities. Besides, the iteration count is required to be not fewer than 1000.

\smallskip
\noindent
\verb|javax.net.ssl| \textit{Classes}. The methods of Java classes \verb|TrustManager|, \verb|HostnameVerifier| and \verb|SSLSocketFactory| in \verb|javax.net.ssl| package provide the SSL/TLS services.  Issues usually happen when developers override the default methods or skip necessary steps to bypass the proper verification.

\subsection{CryptoGuard}
CryptoGuard~\cite{rahaman2019cryptoguard} applies backward and forward program slicing to discover constant sources and configurations causing Java cryptographic API misuses. It has implemented a set of refined slicing algorithms to achieve high precision.

\noindent
\textbf{False Positive Reduction.}
CryptoGuard adopts five refinement insights to remove the language-specific irrelevant elements that cause false positives. During  the analysis process, the state indicators (e.g., \verb|getBytes("UTF-8")|), resource identifiers (e.g., keys of a map),  bookkeeping indices (e.g., size parameters of an array),  contextually incompatible constants, and constants in infeasible paths are removed by refinements conditioned on Jimple, which is an intermediate representation of Soot~\cite{vallee2010soot}.

\noindent
\textbf{Runtime Improvement.}
The most costly parts of the inter-procedural analysis are usually the iterative orthogonal explorations.
CryptoGuard improves the runtime by limiting the orthogonal explorations to depth 1, whereas deeper orthogonal method calls are handled by the refinement insights. 

\subsection{Dataflow Analysis in CryptoGuard and Parfait.}
Parfait supports various static program analyses. An important feature of Parfait that is not present in CryptoGuard~\cite{rahaman2019cryptoguard} is the IFDS analysis framework\footnote{The project Heros~\cite{bodden2012inter} implements the IFDS framework on top of Soot, however, CryptoGuard only uses the FlowAnalysis library in Soot, which does not provide IFDS.}. 

\noindent
\textbf{Dataflow Analysis in CryptoGuard.}
  CryptoGuard achieves dataflow analysis based on Soot's \verb|FlowAnalysis| library.  \verb|FlowAnalysis| includes the intra-procedural dataflow analysis that maintains a flow set and updates it along the dataflow traces. 
  CryptoGuard iteratively runs its intra-procedural analysis for callee and caller methods on the call graph. However, this design can result in re-exploring callee methods multiple times. To reduce complexity, its implementation sets the default depth of the clipping callee method exploration to 1.


\noindent
\textbf{IFDS in Parfait.}
Parfait contains both a classic dataflow analysis and analysis using the IFDS algorithm. 
The IFDS framework reduces the dataflow analysis into a graph reachability problem and performs the analysis by building edges among the data facts (i.e., variables) of certain program points. The reachability can be summarized and queried for the future usage to avoid unnecessary re-analysis as much as possible.  

\noindent
\textbf{Parfait Framework.}
To improve scalability, Parfait offers a layered framework to optimize the ensemble of static program analyses. According to the time cost, the analyses are scheduled from the quickest to the slowest. In this way, more bugs can be found with a lower time overhead.  Specifically, in cryptographic vulnerability detection, we dynamically schedule the analyses into different layers according to the depth of callers. More details are in Section~\ref{sec:implementation}.

\section{Detection Methods and Implementation}
Our detection covers all the misuses shown in Table~\ref{tab:crypto_api_usage}. Two scalability enablers of it are the layered framework of Parfait and the summarization mechanism in IFDS to handle callee methods. 

\subsection{Detection Methods}\label{sec:dmethod}
 The detecting logic is similar to CryptoGuard which maps the cryptographic API misuses to the dataflow analysis problems. In terms of the specific detection methods, there are three groups. 

\noindent
\textbf{Group 1: Inter-procedural Backward Dataflow Analysis.}
This group includes the API misuses determined by constant sources. Specifically, these are APIs in Table~\ref{tab:crypto_api_usage} of Java Class \verb|SecureRandom|, \verb|MessageDigest|, \verb|Cipher|, \verb|KeyStore|, \verb|SecretKeySpec|,  \verb|PBEKeySpec|, \verb|PBEParameterSpec|, and \verb|IvParameterSpec|. We require an inter-procedural backward dataflow analysis to capture the constant sources of the API arguments. We apply different verifying rules to the collected constant sources according to the vulnerability types. The verifying rules include whether it is a constant, whether it is a number less than 1000, or whether it matches some weak algorithms (e.g., "DES").  

\noindent
\textbf{Group 2: Intra-procedural Pattern Matching.}
The vulnerabilities related to \verb|TrustManager|, \verb|HostnameVerifier|, and \verb|SSLSocketFactory| in Table~\ref{tab:crypto_api_usage} belong to this group. These vulnerabilities often happen within one method that is responsible for authentication operations. We find them by the intra-procedural pattern matching. Specifically, for \verb|HostnameVerifier|,  we detect whether the return value of the method \verb|verify| is always ``True'' regardless of the verification. For \verb|TrustManager|, we detect three vulnerable patterns in the \verb|checkClientTrusted| and \verb|checkServerTrusted| methods including 1) missing verification behavior; 2) catching the verification exception without throwing it; 3) missing verification under a certain path. For \verb|SSLSocketFactory|, we perform the intra-procedural pattern matching to check whether the \verb|HostNameVerifier.verify| method is called after the \verb|SSLSocketFactory| instance creation.  

\noindent
\textbf{Group 3: Sanitizer vs. Verifier.}
In cryptography operations, \verb|Random| is not strong enough~\cite{random}. However, it is unreasonable to report every \verb|Random| used in a program as a vulnerability. Therefore, we regard \verb|Random| as a verifier and \verb|SecureRandom| as a sanitizer for the traced arguments in group 1. Accordingly, we only report \verb|Random| in these cryptographic usages.

\subsection{Cryptographic Vulnerability Detection Implementation}\label{sec:implementation}
Supported by Parfait, we implement the inter-procedural flow-, context-, and field-sensitive backward dataflow analysis for cryptographic vulnerability detection. Next, we introduce several specific features of Parfait for scalability and good precision.  

\begin{figure}
    \centering
    \includegraphics[width=0.47\textwidth]{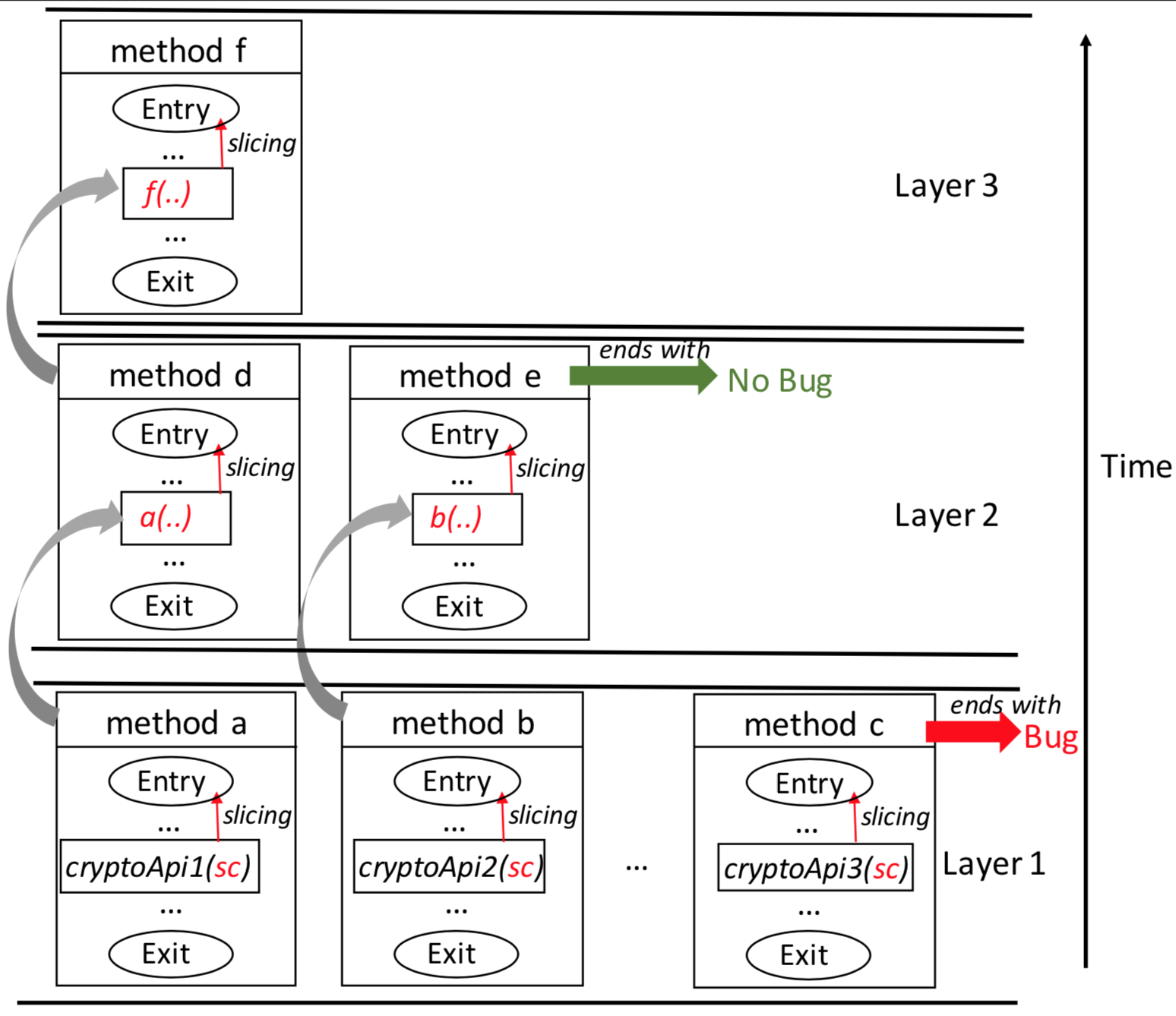}
    \caption{The inter-procedural analysis under Parfait's layered framework. This design is important to achieve the scalability of Parfait.}
    \label{fig:parfait_framework}
\end{figure}

\noindent
\textbf{Layered Scheduler for Caller Methods.}
Parfait optimizes the analysis ensemble to improve scalability. 
Figure~\ref{fig:parfait_framework} demonstrates the backward analyses that are broken down and assigned to different layers. The analyses are scheduled layer by layer. At each layer, the backward analysis ends up at the entry point of the current method with three situations. First, a real bug is verified. Second, the potential bug is sanitized as no bug. Third, further analyses are required in its caller methods. Further analyses will be scheduled at the next layer. In this way, the analysis requiring less time can be performed first.  It also avoids the duplicated parts of two potential vulnerabilities detection traces. 
This layered framework effectively improves the efficiency of finding bugs.

\noindent
\textbf{Flow Functions in IFDS.}
There are several flow functions used to define the analysis. In our cryptographic vulnerability detection, they are:
\begin{itemize}
    \item \verb|flow|: This function specifies the dataflow edges through ordinary non-call instructions. Specifically, it applies to the LLVM instructions \verb|ReturnInst|, \verb|LoadInst|, \verb|StoreInst|, and \verb|BitCastInst|. 
    
    \item \verb|phiFlow|: This function specifies the dataflow edges through the LLVM \verb|phi| instruction.
 
  \item \verb|returnVal|: The function specifies the dataflow edges between the \verb|ReturnInst| of the callee method and its callsite. The summary edges of the callee method are queried at this point to handle the callee method. 

    \item \verb|passArgs|: The function specifies the dataflow edges between the arguments of the callee method and the parameters passed in its callsite.  

    \item \verb|callFlow|: The function handles the dataflow edges regardless of the callee method. Most of the refinements happen here to handle the callee method whose implementation is unavailable.

\end{itemize}

The major differences of these flow functions between the analysis for cryptographic vulnerabilities and taint analysis are the dataflow edges from constants. The cryptographic vulnerability detection discovers the edges flowing out from constants and refines them according to five refinement insights, which does not happen in the taint analysis. Furthermore, cryptography vulnerability detection redefines the default dataflow edges in \verb|callFlow|. More details are in Section~\ref{sec:refinement}. 

\noindent 
\textbf{Summarization for Callee Methods.}
Another design improving the scalability is the summarization mechanism for the callee methods. After a method is explored, the summary edges for it are stored for future usage.  Parfait exhaustively summarizes all methods in advance and queries the summary edges of the callee methods on demand. All the methods are summarized in a bottom-up manner according to the call graph, beginning from leaf methods to their callers. This design guarantees every method is only explored once. Hence, the re-exploration for callee methods is eliminated to avoid complexity explosion.


%


\begin{figure}
    \centering
    \includegraphics[width=.95\linewidth]{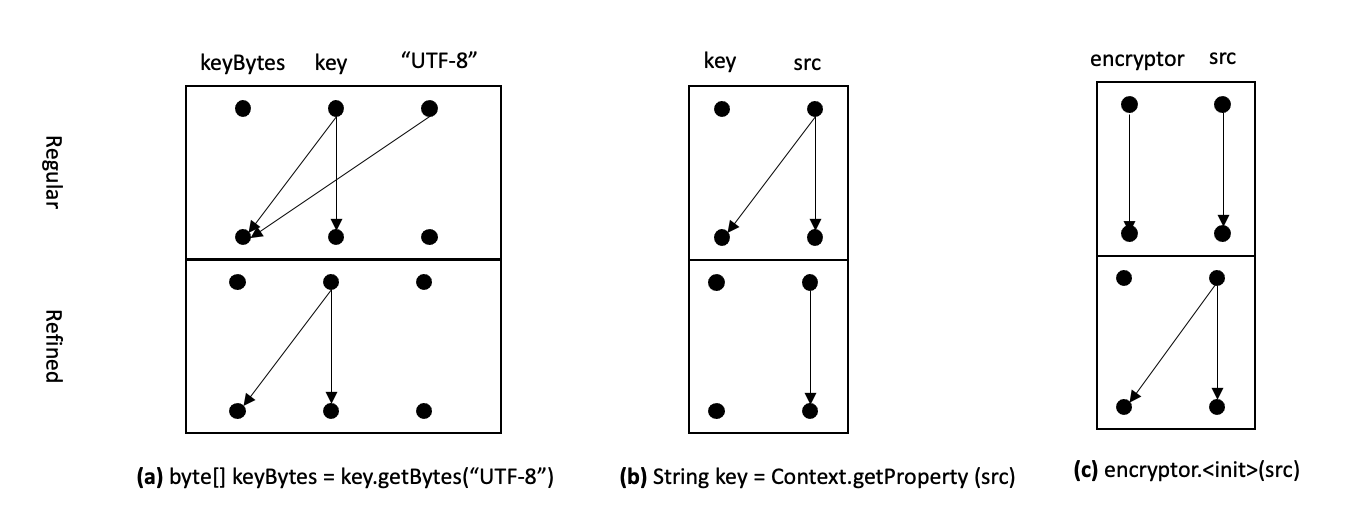}
    \caption{The false-positive reduction refinements represented in IFDS. It shows the dataflow propagating edges for three different types of method calls, 1) a virtual method call with a return value, 2) a static method call with a return value, and 3) a virtual method without a return value. The above ones are the default propagating edges. The bottom ones are the refined propagating edges.}
    \label{fig:heuristics}
\end{figure}

 \subsection{Pseudo-influences and Refined Analysis}\label{sec:refinement}
 
 \textbf{Pseudo-influences.} 
 We use the backward dataflow analysis to capture the constants involved in constructing a security-critical value. When a constant is used to hard-code the security-critical value (e.g., secret key, password), it may cause vulnerability by exposing sensitive information.  However, some constants do not have security impacts on the value, referred to as pseudo-influences. Static analysis is unable to identify them. Reporting all the captured constants as a dangerous source leads to an extremely high false-positive rate. In the work CryptoGuard~\cite{rahaman2019cryptoguard}, the authors summarize five language-specific scenarios that use constants without resulting in hard-coded values. These scenarios include using constants as a state indicator, resource identifier, and bookkeeping indices to retrieve the value.  The contextually incompatible constants, and constants in infeasible paths are also regarded as pseudo-influences.
 
 \smallskip
 \noindent
 \textbf{Refined Dataflow Analysis.}
We refine our dataflow analysis to exclude these pseudo-influences and thus achieve good precision. 
According to the refinement insights from CryptoGuard, we define our pseudo-influence excluding rules in the context of IFDS algorithms and LLVM IR instructions. We select \verb|callFlow| function in our IFDS dataflow analysis to apply the refinement rules. The reason is that most of the pseudo-influences appear as the arguments of a method call. For example, the pseudo-influence \verb|"UTF-8"| is the argument of the method \verb|<String: byte[] getBytes(String)>|. 

In the form of IFDS, we describe the rules with the graph reachability between the data variables given an LLVM instruction. As shown in Fig.~\ref{fig:heuristics}, the data flow edges are refined according to the method signature we obtained from the LLVM instruction. Specifically, there are three types of call instructions. We apply different data flow propagation rules to them. First, if the call instruction has a return value and invoking an instance method that belongs to an object, we change the default data flow propagation edges as described in Fig.~\ref{fig:heuristics} (a). The edge from the argument to the return value is eliminated because the argument is likely to be a pseudo-influences. Second, if the call instruction has a return value and invoking a static method without an associated object, we also eliminate the edge from its argument to the return value to avoid pseudo-influences, as shown in Fig.~\ref{fig:heuristics} (b). Finally, if the call instruction does not have a return value and belongs to an object, we add a data flow edge from its argument to the object holder. Meanwhile, we remove the edge between the object holder itself before and after this call instruction. This allows us to stop tracing the object but tracing the argument that influences the object.   
The example is given in Fig.~\ref{fig:heuristics} (c).



\section{Accuracy Analysis and Real-world Findings}

We have tested our cryptographic vulnerability detection on a comprehensive cryptographic vulnerability benchmark (CryptoAPI-Bench \cite{afrose2019cryptoapi}) to evaluate the precision and recall. To learn its scalability, we further perform experiments by scanning eleven large real-world codebases to obtain the runtime performance. 

 \subsection{Accuracy Analysis on CryptoAPI-Bench}   

\begin{table}
\centering
\caption{Parfait's evaluation results on 158 test cases from CryptoAPI-Bench. We show the numbers of insecure cases, secure cases, reported cases, false positives (FPs) and false negatives (FNs). The 158 test cases include basic cases (intra-procedural), and different inter-procedural cases that require across methods, across classes, field sensitivity, path-sensitivity, and heuristics to handle.}
\label{tab:crypto_bench_results}
\begin{tabular}{|l|c|c|c|c|c|c|c|c|}         
\hline
\textbf{Type}     & Test Cases  & Insecure & Secure & Reported & FPs & FNs & Precision        & Recall           \\ \hline
Basic Cases       & 27           & 24             & 3            & 24             & 0               & 0               & 100\%            & 100\%            \\ \hline
Multiple methods   & 57           & 56             & 1            & 54             & 0               & 2               & 100\%            & 96.43\%          \\ \hline
Multiple Classes    & 23           & 18             & 5            & 18             & 0               & 0               & 100\%            & 100\%            \\ \hline
Field Sensitivity & 19           & 18             & 1            & 18             & 0               & 0               & 100\%            & 100\%            \\ \hline
Path Sensitivity  & 19           & 0              & 19           & 19             & 19              & 0               & 0 \%             & 0 \%             \\ \hline
Heuristics        & 13           & 9              & 4            & 9              & 0               & 0               & 100\%            & 100\%          \\ \hline
Total             & \textbf{158} & \textbf{125}   & \textbf{33}  & \textbf{142}   & \textbf{19}     & \textbf{2}      & \textbf{86.62\%} & \textbf{98.40\%} \\ \hline
\end{tabular}
\end{table}

We have tested Parfait on 158 test cases from CryptoAPI-Bench~\cite{afrose2019cryptoapi}. CryptoAPI-Bench includes various kinds of test units from basic ones to more advanced cases. The basic test cases only require intra-procedural analysis to handle. The advanced cases are inter-procedural ones that require analyses across multiple methods, multiple classes, achieving field sensitivity, and path sensitivity.

 The breakdown numbers are shown in Table~\ref{tab:crypto_bench_results}. The overall precision and recall are 86.62\% and 98.40\%, respectively. All the false positive cases come from path sensitivity cases, which verifies that our tool has achieved high precision for the cases excluding path-sensitive ones. We analyzed several examples to further reveal the details of Parfait cryptographic vulnerability detection and discuss possible improvements.  

\noindent
\textbf{Impact of Refinement Insights.} We demonstrate the impact of our refinement insights by comparing the Parfait cryptographic vulnerability detection with its intermediate version that does not have the refinement strategies.  Table~\ref{tab:impact_heuristics} shows the comparison. Without the refinements, there are 38 false positive cases. Based on our manual analysis, most of the false positives are caused by the pseudo-influences we introduced in Section~\ref{sec:refinement}. The refinement insights successfully reduce all the false positive cases except for the path-sensitive case.

\begin{table}[]
\centering
\caption{False positive reduction derived from applying the refinement insights (RIs). We compare Parfait cryptographic vulnerability detection with its intermediate version without the refinement insights.}
\label{tab:impact_heuristics}
\begin{tabular}{|l|c|c|c|c|}
\hline
\textbf{Type}     & \# of Vulnerabilities & FPs (w/o RIs) & FPs (with RIs) & Reduction     \\ \hline
Basic Cases       & 24                    & 1                    & 0                     & 100\%         \\ \hline
Multiple Methods  & 56                    & 3                    & 0                     & 100\%         \\ \hline
Multiple Classes  & 18                    & 1                    & 0                     & 100\%         \\ \hline
Field Sensitivity & 18                    & 2                    & 0                     & 100\%         \\ \hline
Path Sensitivity  & 0                     & 19                   & 19                    & 0             \\ \hline
Heuristics        & 9                     & 12                   & 0                     & 100\%         \\ \hline
Total             & \textbf{125}          & \textbf{38}          & \textbf{19}           & \textbf{50\%} \\ \hline
\end{tabular}
\end{table}

\subsection{Evaluation on Real World Projects}\label{sec:real-world-findings}

We evaluated our tool on eleven real world codebases. Nine of them are \anonymous{Oracle}{}internal products \anonymous{}{of a large software company} while two are open-source projects Spring-Security\footnote{https://github.com/spring-projects/spring-security} and Orchid\footnote{https://github.com/OrchidTechnologies/orchid}. We select these projects because they are security-relevant and use Java cryptographic APIs. 

\subsubsection{Runtime and Precision}

\begin{figure}

    \centering
    \includegraphics[width=.6\linewidth]{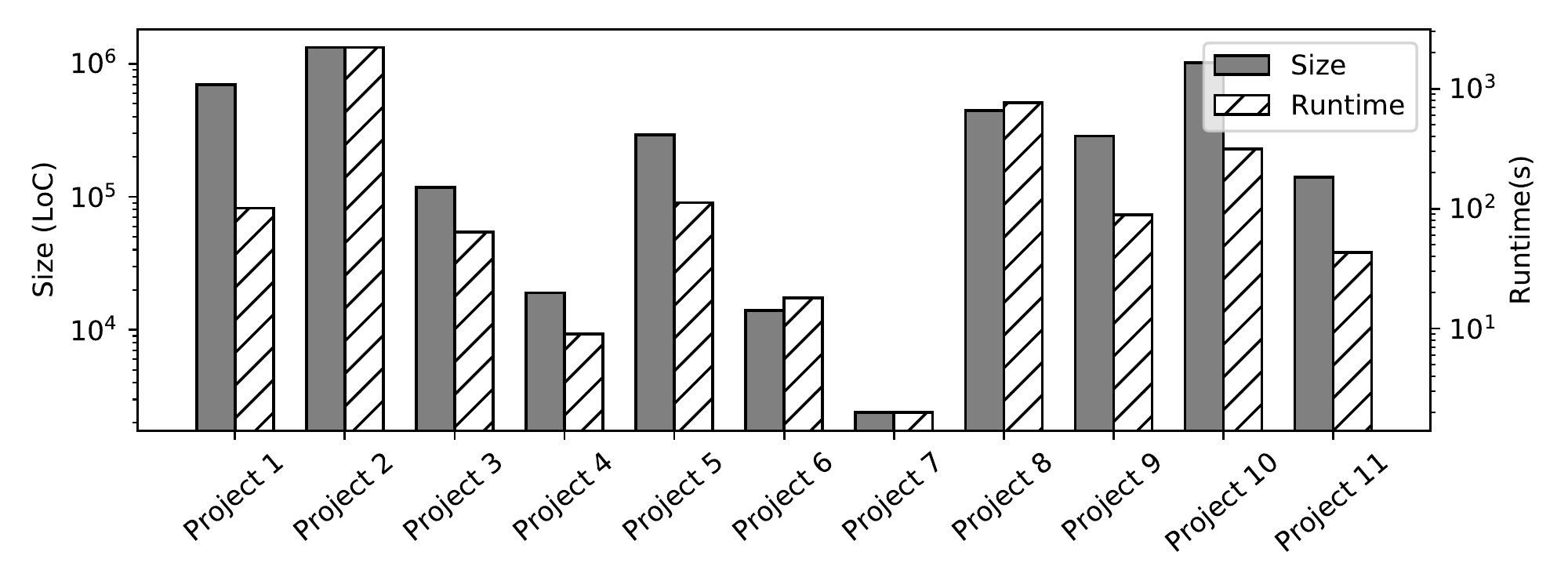}
   
    \caption{Runtime performance of Parfait for screening the eleven real-world codebases. The size shows how many lines of code these codebases have.  }
    \label{fig:runtime}
 
\end{figure}

Scalability is always one of the most important concerns. We list the runtime performance and the size of these scanned projects in Fig.~\ref{fig:runtime}. The project sizes vary from 2K to 1321K. The detection is run on the machine with Intel(R) Xeon(R) CPU E5-2690 v4 @ 2.60GHz, 128G memory, and  Oracle Linux Server release 6.9 operating system. The results show that Parfait achieves excellent scalability. 
The analysis can be finished within 10 minutes for the majority of these projects including those with millions of lines of code (Project 10). 

\begin{figure}
    \centering
    \includegraphics[width=.6\linewidth]{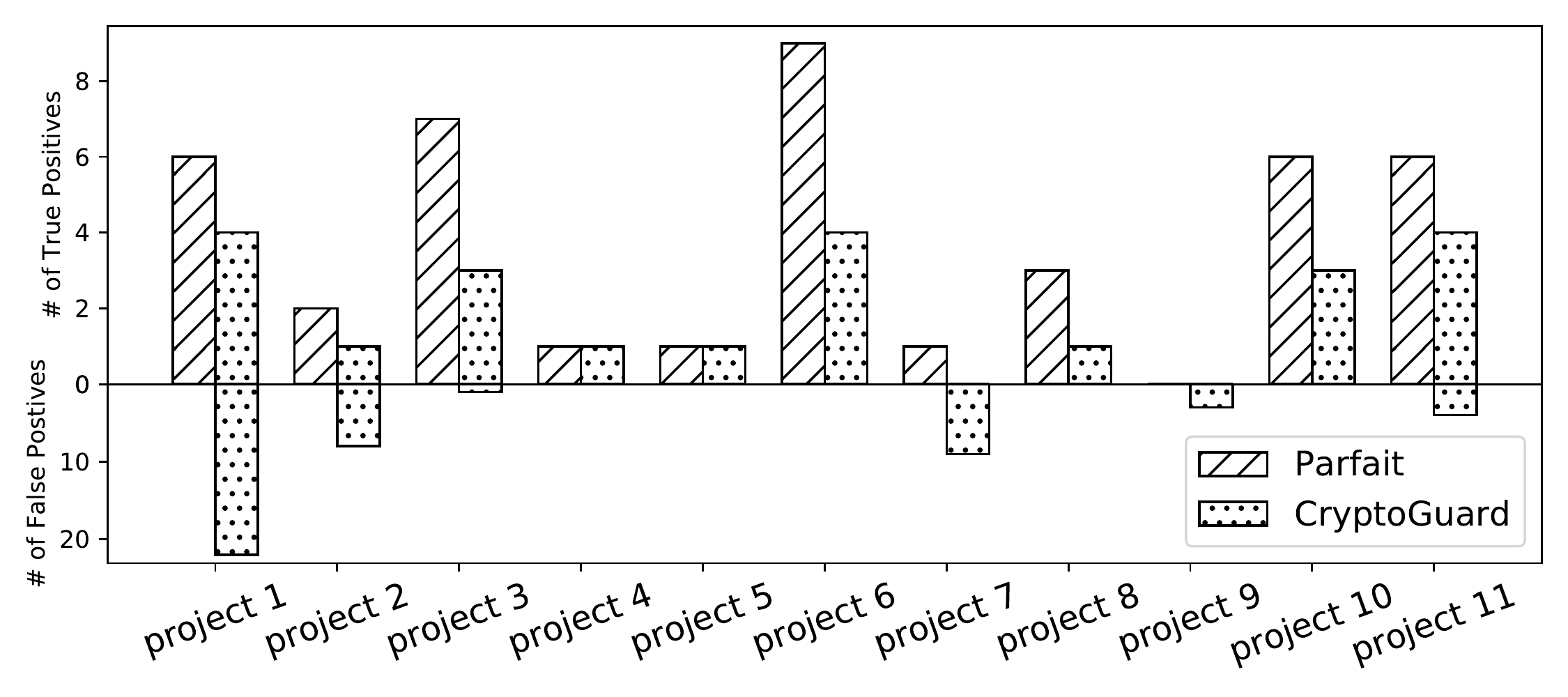}
    \caption{The number of vulnerabilities reported by Parfait and CryptoGuard in the eleven real-world industrial applications. The upper area of the x axis shows the true positive alerts while the bottom area of the x axis shows the false positive alerts.  Nine of them are \anonymous{Oracle}{}internal codebases of \anonymous{}{a large software company}. Two of them are open-source projects.}
    \label{fig:real_world_acc}
    
\end{figure}

Fig.~\ref{fig:real_world_acc} demonstrates the precision results of Parfait and CryptoGuard on the eleven real-world projects. Compared with CryptoGuard, Parfait successfully identified more true positive cases with fewer false positives. Parfait reported 42 vulnerabilities and all of them are manually verified as true positives. The precision reaches 100\%. We show several real-world vulnerabilities found by Parfait in Section~\ref{sec:findings} CryptoGuard reported 69 vulnerabilities. However, there are 47 false positives among them. The precision is 31.88\%. We noticed that all the false positive cases of CryptoGuard are caused by the same issue, that is, how CryptoGuard detects weak Pseudo-random Number Generator (PRNG) vulnerabilities. We noticed that all the false positives of CryptoGuard are caused by the same issue, that is, how CryptoGuard identifies weak PRNG cases. We will discuss it in the comparison between CryptoGuard and Parfait. 
 
\smallskip
\noindent
\textbf{Comparison with CryptoGuard.} As we introduced, CryptoGuard and Parfait leverage identical refined dataflow analysis at a high level to detect cryptographic vulnerabilities. Here, we analyze the differences between them in detection results.  

\begin{figure}

    \centering
    \includegraphics[width=.6\linewidth]{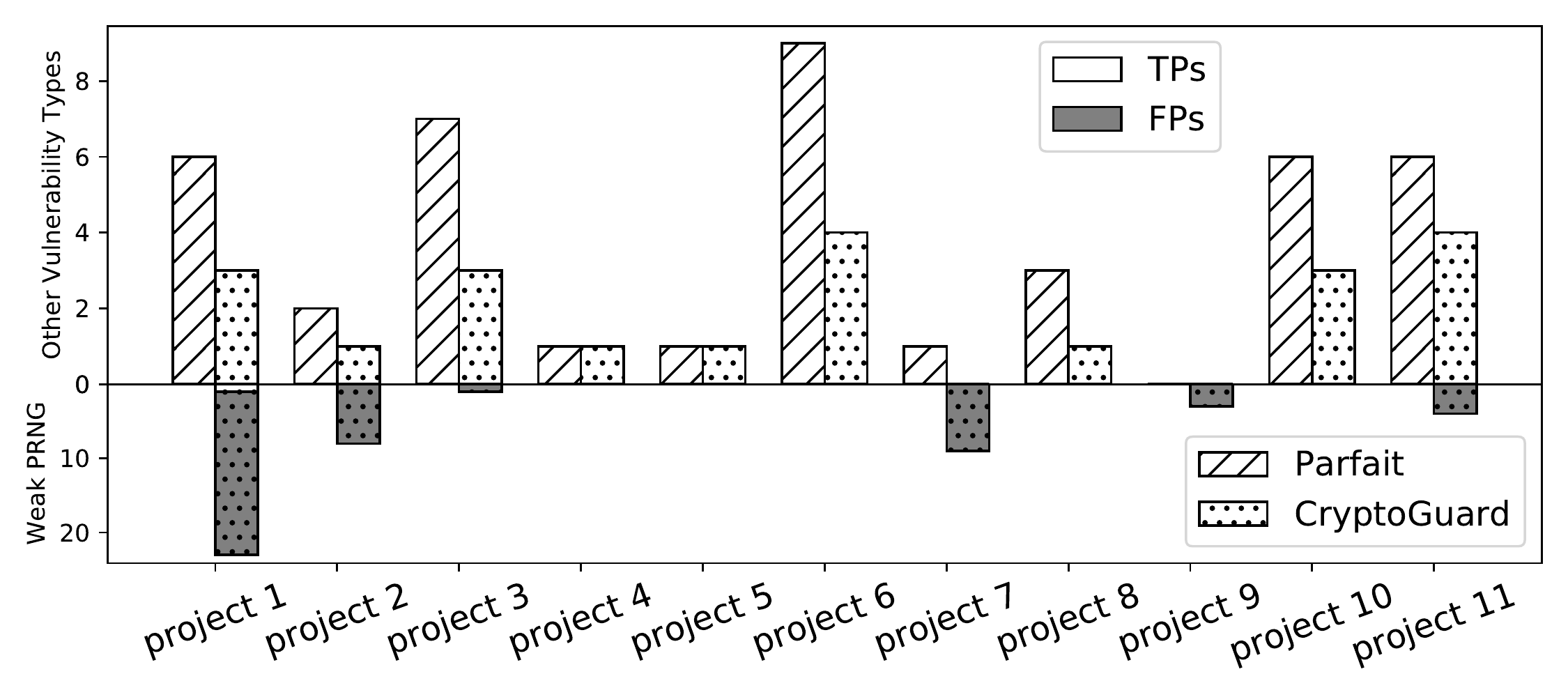}
   
    \caption{The number of vulnerabilities reported by Parfait and CryptoGuard in the eleven real-world industrial applications. We break them down into the weak PRNG vulnerabilities and the other vulnerability types. Parfait reported 0 weak PRNG vulnerability. CryptoGuard reported 48 weak PRNG vulnerabilities while only 1 of them is a true positive case. Parfait and CryptoGuard both achieves 100\% precision in the other vulnerability types excluding the weak PRNG cases.}
    \label{fig:cryptoguard}
 
\end{figure}

\smallskip
\noindent
\textit{Detection for Weak PRNG.}
A major difference between Parfait and CryptoGuard is the way they identify weak PRNG vulnerabilities. After manual analysis, we noticed that all the false positives of CryptoGuard shown in Fig.~\ref{fig:real_world_acc} are weak PRNG cases. To make it more clear, we break down the reported cases into weak PRNG cases and other types of vulnerabilities, as shown in Fig.~\ref{fig:cryptoguard}. Overall, there are 48 weak PRNG vulnerabilities and 21 other types of vulnerabilities reported by CryptoGuard. Among the 48 weak PRNG cases, only 1 of them is verified as a true positive case.  As a contrast, Parfait reported 0 weak PRNG case, which indicates that Parfait missed at least 1 weak PRNG vulnerability.  This suggests that CryptoGuard tends to have a more conservative approximation on weak PRNG vulnerability detection while Parfait reports this type of vulnerability in a more precise approximation.

Listing~\ref{lst:weak_prng} shows a false positive weak PRNG identified by CryptoGuard. The Java class \verb|Random| is not strong enough, therefore,  an alternative class \verb|SecureRandom| that is cryptographically strong is recommended to use. However, our manual verification confirmed that the \verb|Random| instance is not used in a security or cryptographic context. Hence, we consider it is a false positive as there is no impact on security. 
CryptoGuard performs an exhaustive search in the codebase to report every \verb|Random| usage regardless of the context. Hence, there are many false positives. On the opposite, Parfait applies a more strict criterion for alerting this type of vulnerability. Only when the \verb|Random| instance is passed to the cryptographic APIs covered in Table~\ref{tab:crypto_api_usage}, will it be reported as a weak PRNG case. However, this may miss some cases due to the limited coverage of the cryptographic APIs. 
It is difficult to accurately determine whether a \verb|Random| instance is used for cryptographic purposes. Identifying more vulnerable usage patterns and involved cryptographic APIs for this type of vulnerability can be future work. To extend the current detection criteria, Parfait provides the flexibility for users to change the sinks, sources, sanitizers, and verifiers of the dataflow analysis through configuration, which makes customizing the vulnerability detection rules easy. 

\begin{lstlisting}[language=Java,escapechar=\%,caption={A reported weak PRNG vulnerability that is a false positive},captionpos=b,label={lst:weak_prng}]
Random random = new Random();
int rnumber = random.nextInt();

\end{lstlisting}

\noindent
\textit{Exploration Depth for Callee Methods.} Another difference between Parfait and CryptoGuard is the exploration depth for callee methods when performing interprocedural analysis. The interprocedural dataflow analysis requires exploring the encountered callee methods. When meeting the recursive callee methods or the callee stack is too deep, the analysis needs to clip the call graph. CryptoGuard allows users to configure the exploration callee stack depth. To make the analysis fast, CryptoGuard set the default callee stack depth as 1. Parfait deals with this problem by summarization mechanism (see details in Section~\ref{sec:implementation}). This design avoids clipping the callee stack, however, the price is that the summarization becomes the most costly part.  To make the summarization as a one-time cost, it is performed separately in advanced and stored for queries when encountering a callee method in the dataflow analysis. 
In Fig.~\ref{fig:real_world_acc}, we observe that CryptoGuard missed 21 cases that have been reported by Parfait. This might be attributed to the limited default callee stack depth that CryptoGuard explores. It can be improved by setting a larger value of the callee stack depth.

\noindent 
\textit{Application Perspective vs. Library Perspective.} 
Parfait differs from CryptoGuard in the vulnerability definitions in some situations. An example is given in Listing~\ref{lst:no_caller} in the Appendix. If the potentially vulnerable method is not called in the scanned codebase, the concerned field variable is left undetermined and then Parfait considers it as a non-vulnerable case. However, CryptoGuard applies a forward slicing for this field variable to find out the possible assignments in the initialization. If a constant is assigned in the initialization, CryptoGuard still considers it as a vulnerability. If the detected issues are in applications, Parfait's design is superior because it avoids overestimating the vulnerabilities. If they are in libraries, CryptoGuard's design is better as it discovers the potential buggy method although they may not be called yet. 

\subsubsection{Real-world Findings}\label{sec:findings}

We have reported the detected vulnerabilities to corresponding developers. In terms of the open-source projects, we further find that the vulnerabilities are either in their non-production (development) mode or fixed in their latest versions. 
  We show several real-world detected cases below. 



\begin{lstlisting}[language=Java,escapechar=\%,caption={A real-world vulnerability about using constant salt and insufficient iteration count (We modified the code to make the codebase unidentifiable.)},captionpos=b,label={lst:real_case2}]
public class DesEncrypter{
%\includegraphics[scale=.015]{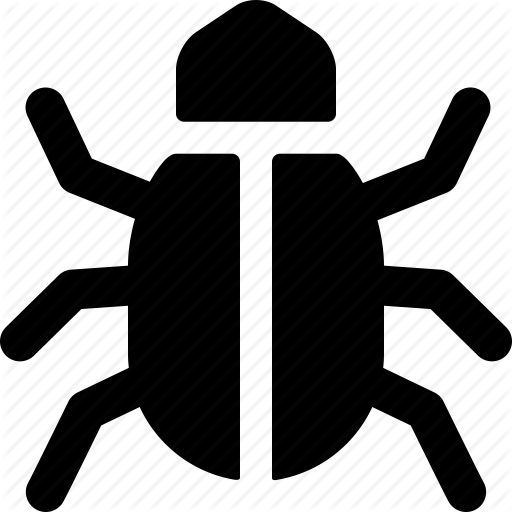}%  private byte[] %\underline{salt}% = { (byte) 0xC9, (byte) 0xDB, (byte) 0xA3, (byte) 0x52, (byte) 0x56, (byte) 0x35, (byte) 0xE8, (byte) 0xB0};
%\includegraphics[scale=.015]{bug.png}%  private int %\underline{iterationCount}% = 20;
    public DesEncrypter(final String passPhrase){
        initDesEncrypter(passPhrase);}
    private void initDesEncrypter(final String passPhrase){
    ...
    AlgorithmParameterSpec paramSpec = new PBEParameterSpec(%\underline{salt},\underline{iterationCount}%);}}
\end{lstlisting}

Listing~\ref{lst:real_case2} shows vulnerabilities of using constant salt and insufficient iteration count as PBE parameters. This case represents the most common vulnerable pattern of the sensitive cryptographic materials (e.g., passwords, salts, IVs, etc) to be hard-coded in the initialization.  

\begin{lstlisting}[language=Java,escapechar=\%,caption={A real-world vulnerability about insufficient entropy salts},captionpos=b,label={lst:real_iterative_salt}]
public String padding_salts(String salts){
    StringBuffer sb = new StringBuffer();
    for(%\underline{int i=salts.getBytes().length; i<16; i++}%){
 %\includegraphics[scale=.015]{bug.png}%     %\underline{sb}%.append((byte)%\underline{i}%&0xfe)}
    String padded_salts = %\underline{salts}%+%\underline{sb}%.toString();
    return %\underline{padded\_salts}%;}
\end{lstlisting}

Listing~\ref{lst:real_iterative_salt} is a noteworthy real-world example. It introduces a vulnerability of using salts with insufficient entropy. When a random salt is iteratively assigned by the same variable, its value space is reduced significantly and hence makes the exhaustive attack feasible. Our analysis reports a constant number 16 at Line 3 involved in the construction of the salts. However, to accurately capture the insufficient entropy issue, symbolic execution is required.


\begin{lstlisting}[language=Java,escapechar=\%,caption={An example from CVE-2019-3795},captionpos=b,label={lst:real_casel}]
public SecureRandom getObject() throws Exception{
    SecureRandom rnd = SecureRandom.getInstance(algorithm);
    if(seed != null){
%\includegraphics[scale=.015]{bug.png}%     byte[] seedBytes = FileCopyUtils.copyToByteArray(seed.InputStream());
        rnd.setSeed(%\underline{seedBytes}%); //manual seeding
    }else{
        rnd.nextBytes(new byte[1]) //self-seeding
    }}
\end{lstlisting}


Listing~\ref{lst:real_casel} shows a detected vulnerability in the open-source project Spring Security, disclosed as CVE-2019-3795~\cite{cve_case}. This vulnerability appears in Spring Security versions 4.2.x before 4.2.12, 5.0.x before 5.0.12, and 5.1.x before 5.1.5. Although not involving a hard-coded seed, the \verb|SecureRandom| instance relies on an unreliable  \verb|InputStream| at Line 4 as the \verb|seed|.  Inspired by this real-world vulnerability, we apply a more strict rule for \verb|SecureRandom.setSeed| to avoid unreliable seeding. Only self-seeding and manual seeding by the method \verb|SecureRandom.generateSeed()| are considered as secure. A self-seeding (secure) will be automatically enforced if the API \verb|nextBytes| is called immediately after the \verb|SecureRandom| instantiation~\cite{securerandom}.

\begin{lstlisting}[language=Java,escapechar=\%,caption={A real-world false positive case about TrustManager},captionpos=b,label={lst:false_positive1_real}]
public void checkClientTrusted(X509Certificate[] certs, String authType) throws CertificateException{
%\includegraphics[scale=.015]{bug.png}%   throw new UnsupportedOperationException("checkclientTrusted is unsupported in "+ this.getClass().getName());}
\end{lstlisting}

Listing~\ref{lst:false_positive1_real} shows a reported case for bypassing certificate verification. This case disables the certificate verification by simply throwing the \verb|UnsupportedOperationException| for all certificates. This misuse, matching a vulnerable pattern, was reported, however it is not enabled in the production code path, and hence not exploitable or requiring any remediation. 

\subsection{Discussion} 

We discuss the potential improvement and limitations of Parfait.

\smallskip
\noindent
\textbf{Potential Improvement.}
There are two potential improvements to fix the false-negative cases. First, a false negative could be caused by missing the summarization for \verb|clinit| method. An example is shown in Listing~\ref{lst:false_negative1} in the Appendix. This deficiency is derived from the fact that \verb|clinit| has not appeared in Parfait's call graph. A fix for this issue could be updating the call graph construction to cover the \verb|clinit| of every class. Second, a false-negative case shown in Listing~\ref{lst:false_negative2} is caused by incompatible types between the captured source (i.e., \verb|String|) and the sensitive argument (i.e., \verb|int|). This corner case can be improved by checking the type compatibility through the type casting in Java language. 

\smallskip
\noindent
\textbf{Limitations.}
Our cryptographic vulnerability detection still has limitations with handling path-sensitive cases and pointer issues. We show a path-sensitive false-positive case in Listing~\ref{lst:false_positive1} in the Appendix.  Furthermore, another potential cause for false positives could be pointer issues. Due to the limitation of static analysis, there may be over-approximation in our call graph construction, which leads to potential false positives. However, path-sensitivity and pointer precision are too costly, in our experience, for large codebases.  Our analysis is designed to scan large-scale industrial projects, therefore we accept the trade-off for better overall performance.



\section{Conclusion and Future Work}

We have implemented a precise and scalable cryptographic vulnerability detection in the framework of Parfait. Leveraging the refinement insights from CryptoGuard, our detection reproduced the high precision results (few or no false positives) achieved by CryptoGuard. Experiments show 100\% precision for eleven real-world large-scale projects and CryptoAPI-Bench, excluding the path-sensitivity cases.  Our cryptographic vulnerability detection benefited from the IFDS and layered framework of Parfait to achieves good runtime performance for large-scale codebases. The runtime for these eleven large-scale codebases ranges from 2 seconds to 36 minutes. Ten of them can be screened within 10 minutes. 

We leverage the backward dataflow analysis for our cryptographic vulnerability detection in Parfait. For future improvement, there are still some remained cases that require other techniques, such as forward dataflow analysis, symbolic execution, to handle. 
Besides, how to improve the detection accuracy of Weak PRNG vulnerabilities by identifying their context is also an interesting future direction. 


\section{Acknowledgement}
The Virginia Tech authors have been supported by the National Science Foundation under Grant No. CNS-1929701.

\bibliographystyle{ACM-Reference-Format}
\bibliography{main}


\appendix

\section{A step-by-step Illustration of Our IFDS analysis}
\begin{figure}
    \centering
    \includegraphics[width=\linewidth]{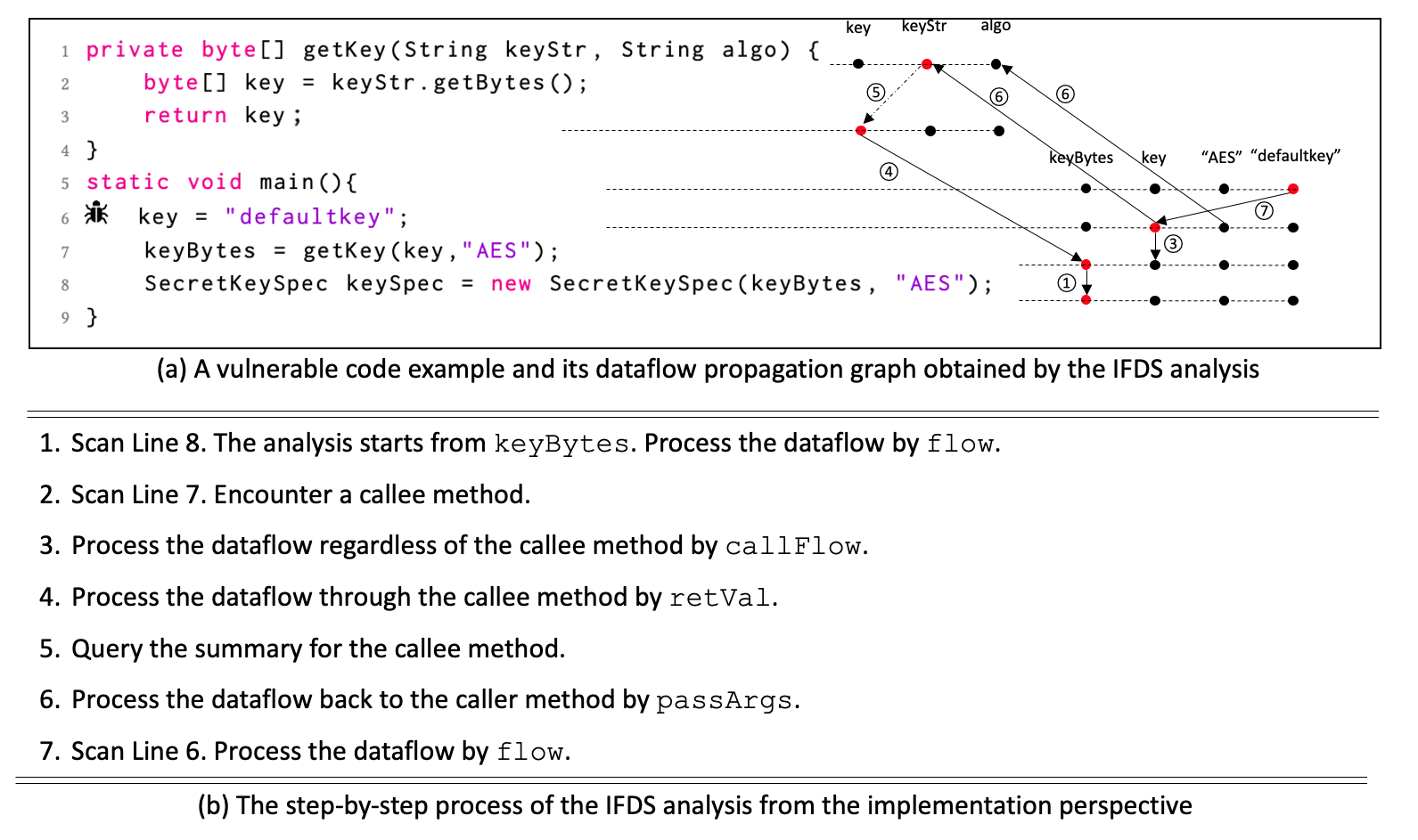}
    \caption{A step-by-step breakdown of A vulnerability detected by our IFDS analysis implementation. (a) shows a vulnerable code snippet with the captured dataflow propagation graph by IFDS analysis. The right side of (a) is a dataflow propagation graph obtained by IFDS analysis. At each program line, there are several dots representing a data fact (variable) at this program point. An edge from dot \lstinline|v1| to dot \lstinline|v2| means there is a dataflow edge from \lstinline|v1| to \lstinline|v2|. The numbering in circles corresponds to the steps in (b), which process the dataflow and draw an edge in the graph.  The red dots form a detected dataflow path from the insecure constant to the targeted cryptographic API. (b) shows the steps of our IFDS analysis of (a) from implementation perspective. \lstinline|flow|, \lstinline|retVal|, \lstinline|callFlow|, etc. are the flow functions defined in Section~\ref{sec:implementation}.}
    \label{fig:steps}
\end{figure}

We give a step-by-step breakdown to show how a vulnerability is captured by our IFDS analysis implementation in Fig.~\ref{fig:steps}. Fig.~\ref{fig:steps} (a) gives a simple example of a detected vulnerability. The analysis starts from Line 8 in the code snippet. A constant ``defaultkey'' at Line 6 is captured by our analysis. The right part shows how the constant ``defaultkey'' is connected to the variable \verb|keyBytes| by dataflow. Fig.~\ref{fig:steps} (b) is the step-by-step process to illustrate how the dataflow propagation is handled by the flow functions (see Section~\ref{sec:implementation}) of our IFDS analysis implementation.  

\section{Ordinary Iterative Analysis vs. IFDS analysis}
\begin{figure}
    \centering
    \includegraphics[width=0.46\textwidth]{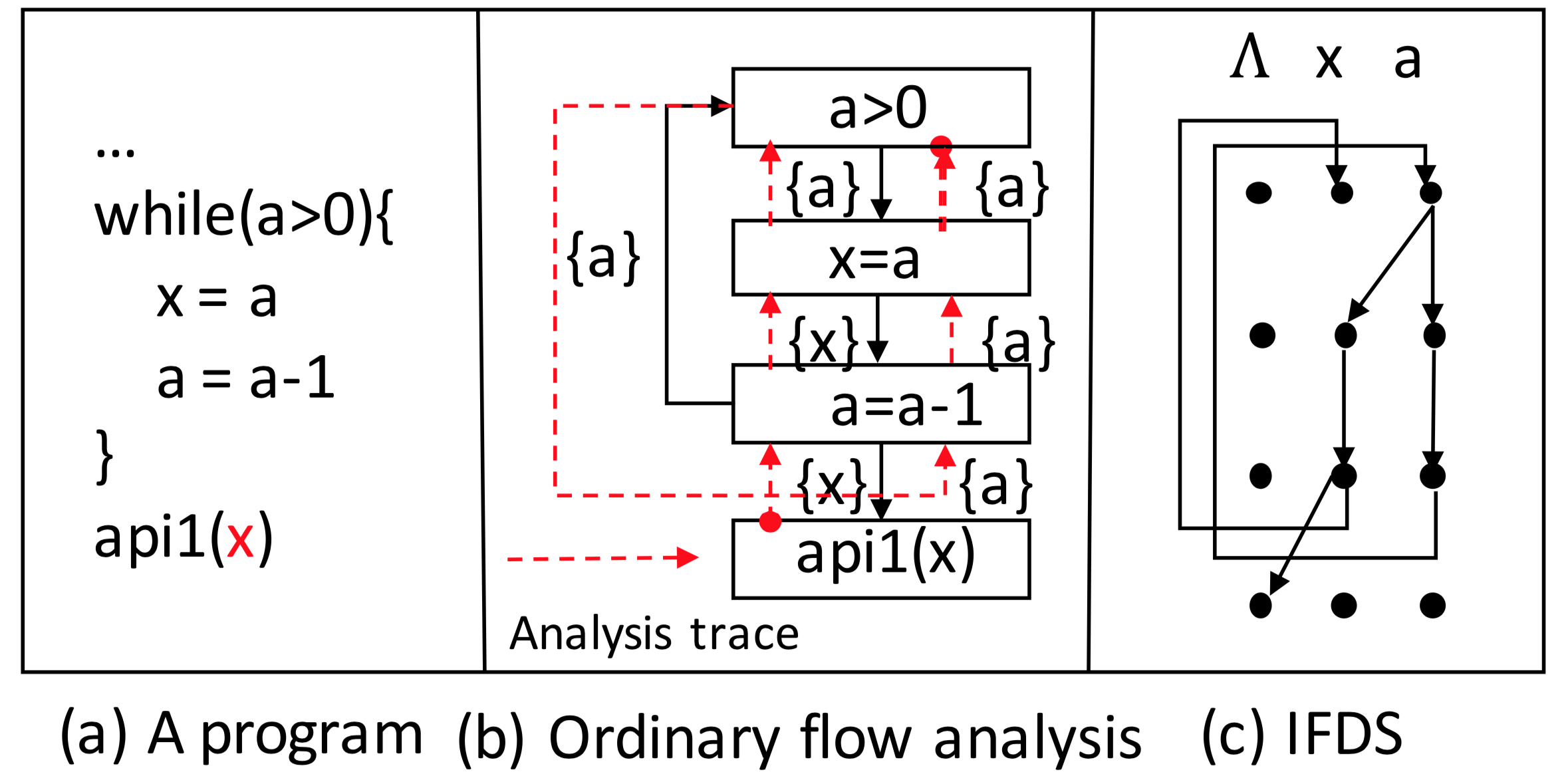}
    \caption{The comparison of the ordinary iterative analysis and IFDS analysis. (a) is a code snippet. (b) shows the ordinary flowset based analysis by collecting and updating a flow set. The code blocks with black edges in (b) represent the control flow graph of (a). The bracket (i.e., \{ \}) between the code blocks represents the flowset at that program point. The flowset keeps track of all the data facts (variables) that can propagates to the entry point of our backward dataflow analysis. (c) shows the dataflow propagation graph obtained by IFDS analysis which builds edges and then summarizes edges during the analysis. $\Lambda$ represents the empty set the backward analysis starts from. Here, we use it as an alert identification. If a dangerous source (e.g., hardcoded key) is connected to $\Lambda$, we will identify it as a vulnerability.  }
    \label{fig:ifds}
\end{figure}

Fig.~\ref{fig:ifds} shows the difference between an ordinary iterative analysis and an IFDS analysis. Fig.~\ref{fig:ifds} (a) is a code snippet. Fig.~\ref{fig:ifds} (b) and (c) are the diagrams of an ordinary dataflow analysis and an IFDS analysis, respectively. As shown in Figure~\ref{fig:ifds}, the ordinary analysis maintain a flowset during the interative analysis. The IFDS framework reduces the analysis as a graph reachability problem and guarantees that the analysis can be finished in polynomial time.

\section{False positive cases in CryptoAPI-Bench}
\begin{lstlisting}[language=Java,escapechar=\%,caption={A false positive caused by path sensitivity},captionpos=b,label={lst:false_positive1}]
    String %\underline{defaultKey}% = %\underline{"defaultkey"}%;
    int choice = 2;
    byte[] %\underline{keyBytes}% = %\underline{defaultKey}%.getBytes();
    //keyBytes-->key material after phiFLow
    if(choice>1){
        //nothing-->key material
        SecureRandom random = new SecureRandom();
        %\underline{keyBytes}% = String.valueOf(random.ints()).getBytes();
    }
    %\underline{keyBytes}% = Arrays.copyOf(%\underline{keyBytes}%,16);
    SecretKeySpec keySpec = new SecretKeySpec(%\underline{keyBytes}%, "AES");
\end{lstlisting}

\begin{lstlisting}[language=Java,escapechar=\%,caption={A false negative case caused due to type matching},captionpos=b,label={lst:false_negative2}]
public class LessThan1000IterationPBEABICase2 {
%\includegraphics[scale=.015]{bug.png}%  public static final String DEFAULT_COUNT = "20";
    private static char[] COUNT;
    private static char[] count;
    public static void main(){ //Bug condition: "20"<1000?
        LessThan1000IterationPBEABICase2 lt = new LessThan1000IterationPBEABICase2();
        go2();    //"20"-->PBE iteration
        go3();    //this.COUNT-->PBE iteration
        lt.key2(); //this.count-->PBE iteration
    }
    private static void go2(){
        COUNT = DEFAULT_COUNT.toCharArray();
    }
    private static void go3(){
        count = COUNT;
    }
    public void key2(){ //this.count-->PBE iteration
        ...
        pbeParamSpec = new PBEParameterSpec(salt, %\underline{Integer.parseInt(String.valueOf(count))}%);
    }
}


\end{lstlisting}

\begin{lstlisting}[language=Java,escapechar=\%,caption={A test cases considered non-vulnerable by Parfait but vulnerable by CryptoGuard. The backward analysis in Parfait terminates at Line 11 and leaves this.crypto.defaultKey as a variable due to no caller of this method.  },captionpos=b,label={lst:no_caller}]
public class PredictableCryptographicKeyABSCase1 {
    Crypto crypto;
    public PredictableCryptographicKeyABSCase1() throws Exception {
        String passKey = PredictableCryptographicKeyABSCase1.getKey("pass.key");
        if(passKey == null) {
            crypto = new Crypto("defaultkey");
        }
%\includegraphics[scale=.015]{bug.png}%      crypto = new Crypto(passKey);
    }
    //this.crypto.defaultKey-->secret key; no caller for encryptPass, terminate
    public byte[] encryptPass(String pass, String src) throws Exception {
        String %\underline{keyStr}% = PredictableCryptographicKeyABSCase1.getKey(src);
        return crypto.method1(pass, keyStr); 
        //keyStr-->secret key; this.crypto.defaultKey-->secret key
    }
    public static String getKey(String s) {
        return System.getProperty(s);
    }
}
class Crypto {
    Cipher cipher;
    String algoSpec = "AES/CBC/PKCS5Padding";
    String algo = "AES";
    String defaultKey;
    public Crypto(String defkey) throws NoSuchPaddingException, NoSuchAlgorithmException {
        cipher = Cipher.getInstance(algoSpec);
        defaultKey = defkey;
    }
    //key-->secret key; this.defaultKey-->secret key
    public byte[] method1(String txt, String %\underline{key}%) throws UnsupportedEncodingException, InvalidKeyException, BadPaddingException, IllegalBlockSizeException {
        if(key.isEmpty()){
            %\underline{key}% = %\underline{defaultKey}%;
        }
        byte[] %\underline{keyBytes}% = %\underline{key}%.getBytes("UTF-8");
        byte [] txtBytes = txt.getBytes();
        %\underline{keyBytes}% = Arrays.copyOf(%\underline{keyBytes}%,16);
        SecretKeySpec keySpec = new SecretKeySpec(%\underline{keyBytes}%,algo); //A potential bug
        cipher.init(Cipher.ENCRYPT_MODE,keySpec);
        return cipher.doFinal(txtBytes);
    }
}
\end{lstlisting}

\begin{lstlisting}[language=Java,escapechar=\%,caption={A false negative case caused due to the summarization},captionpos=b,label={lst:false_negative1}]
   public class PredictablePBEPasswordABICase2 {
%\includegraphics[scale=.015]{bug.png}%  public static String KEY = "sagar";
    public static char [] %\underline{DEFAULT\_ENCRYPT\_KEY}% = KEY.toCharArray(); //"sagar"-->this.DEFAULT_ENCRYPT_KEY happens in clinit
    private static char[] encryptKey;
    ...
    public static void main(String [] args) { //this.DEFAULT_ENCRYPT_KEY-->PBE password
        ...
    }
}
\end{lstlisting}

\end{document}